\documentclass{iau}
\usepackage{graphicx}
\usepackage{academicons}
\usepackage[colorlinks=true]{hyperref}
\usepackage{xcolor}

\title[Dark sky tourism for development] 
{Dark sky tourism and sustainable development in Namibia}

\author[Hannah S. Dalgleish \etal]  
{Hannah S. Dalgleish$^{1,2}$,
Getachew M. Mengistie$^1$, Michael Backes$^{1,3}$, Garret Cotter$^2$, \and Eli K. Kasai$^1$}

\affiliation{$^1$Dept. of Physics, University of Namibia, Pionierspark, Windhoek, Namibia \\ 
[\affilskip]
$^2$Dept. of Physics, University of Oxford, Keble Rd, Oxford, OX1 3RH, UK \\
$^3$Centre for Space Research, North-West
University, Potchefstroom, South Africa \\
email: {\tt {\color{blue}{\href{mailto:hannah.dalgleish@physics.ox.ac.uk}{hannah.dalgleish@physics.ox.ac.uk}}}} \\
}

\pubyear{2020}
\volume{367}  
\pagerange{119--126}
\jname{Education and Heritage in the era of Big Data in Astronomy}
\editors{R.M. Ros, B. Garcia, S. Gullberg, J. Moldon \& P. Rojo, eds.}
\begin{document}

\maketitle

\begin{abstract}
Namibia is world-renowned for its incredibly dark skies by the astronomy community, and yet, the country is not well recognised as a dark sky destination by tourists and travellers. Forged by a collaboration between the Universities of Oxford and Namibia, together we are using astronomy as a means for capacity-building and sustainable socio-economic growth via educating tour guides and promoting dark sky tourism to relevant stakeholders.
\end{abstract}

\firstsection 

\section{Introduction}

\begin{figure}[ht]
\begin{center}
 \includegraphics[width=4.5in]{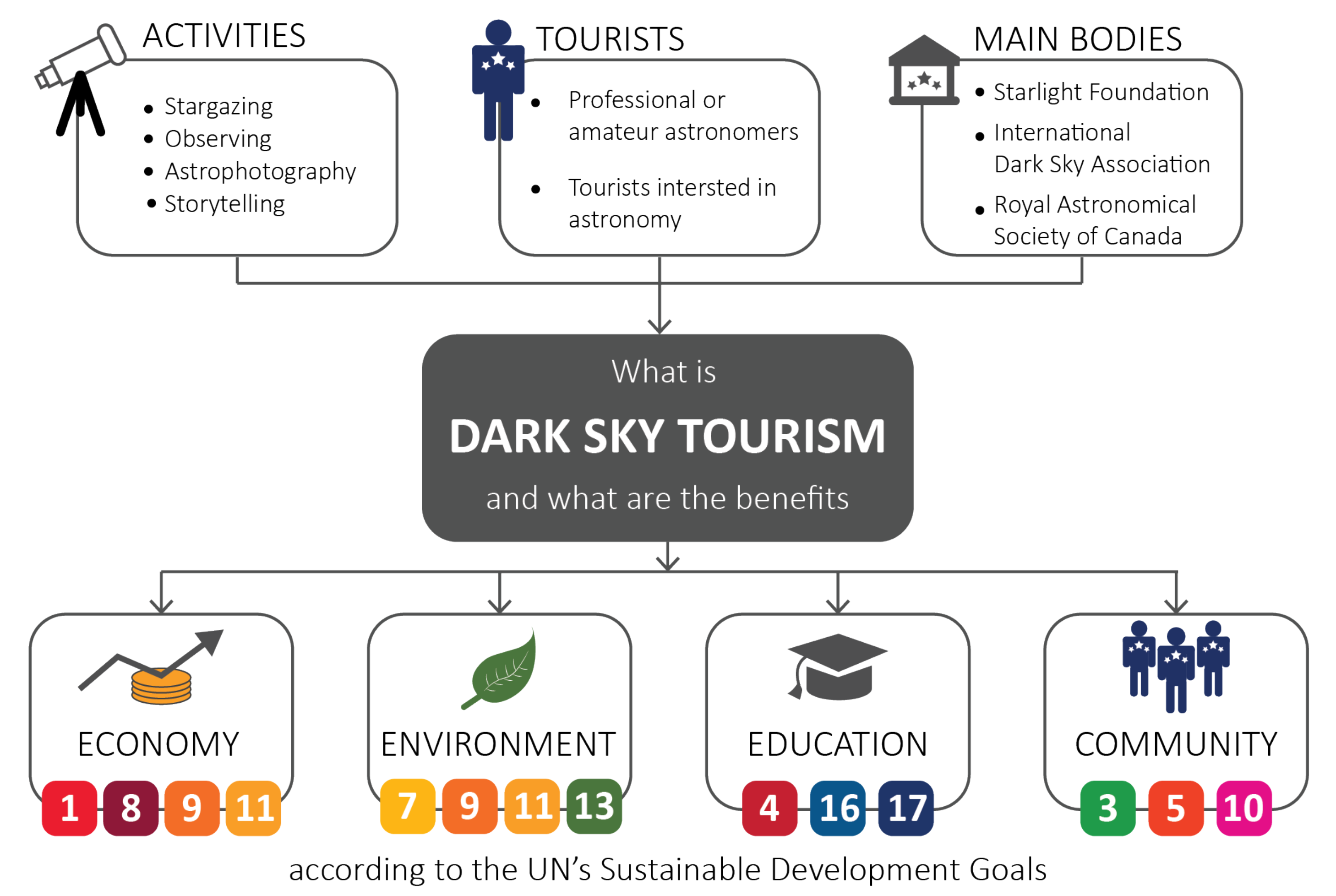} 
 \caption{Dark sky tourism and its relationship with the SDGs (\cite[Dalgleish \& Bjelajac 2021]{Dalgleish21}).}
   \label{fig1}
\end{center}
\end{figure}

Dark sky tourism (DST) attracts visitors to remote, unlit areas to observe celestial objects. Stargazing activities are carried out aided (with binoculars or telescopes) or unaided (with the naked-eye) and can be accompanied by other activities like astrophotography or storytelling. DST has been found to further many of the UN's seventeen Sustainable Development Goals (SDGs), which can help serve as a guide to implement dark sky experiences in a sustainable way (\cite[Dalgleish \& Bjelajac 2021]{Dalgleish21}; Fig.\,\ref{fig1}). 

DST contributes to the SDGs under all branches of sustainability. Economically, DST can generate significant income, providing jobs and extending tourism activity into off-peak times (\cite[Mitchell \& Gallaway 2019]{Mitchell19}; SDG 8---decent work and economic growth). Environmentally, the minimisation of artificial night at light prevents interference with freshwater, marine, and terrestrial wildlife (\cite[Davies \& Smyth 2017]{Davies17}; SDGs 14 and 15---life below water and life on land). Socially, DST presents educational opportunities for tourists and local residents, covering topics from astrophysics and light pollution to indigenous knowledge (\cite[Blundell et al. 2020]{Blundell20}; SDG 4---quality education); and empowers women in rural, underprivileged areas (see e.g. the {\tt {\color{blue}\href{https://astrostays.com}{Astrostays}}} project; SDGs 5 and 10---gender equality and reduced inequalities). Stargazing also promotes health, well-being and connectedness with nature (\cite[Bell et al. 2014]{Bell14}; SDG 3---good health and well-being).

\section{Dark sky tourism in Namibia}

The second least densely populated country in the world, Namibia has minimal light pollution and is therefore very well-suited to dark sky experiences. Namibia is vast with a wide array of climates; some areas (e.g. Sossusvlei) rarely experience cloud cover year-round, and thus, clear and dark skies can be found even during the wet (summer) season. A few lodges and ``astrofarms'' already take advantage of the country's pristine skies, which are especially attractive to amateur astronomers and astrophotographers. Africa's first International Dark Sky Reserve can also be found in Namibia, at the NamibRand Nature Reserve. Thus, there is ample opportunity to extend and promote dark sky activities across the country, especially across wider tourist demographics. 

In order to grow dark sky tourism sustainably, we have been working from both a bottom-up and top-down approach. For the former, we are developing a course comprising five main sections: (1) our place in the Universe, (2) astrophysics research in Namibia, (3) indigenous Namibian star lore, (4) practical astronomy, and (5) light pollution and sustainability. We will be delivering the course to Namibian tour guides in 2021, while ensuring that the content is adaptable for use by similar projects in other countries. We are also exploring options for delivering the course online. At the same time, we are working with the Ministry of Environment, Forestry, and Tourism, as well as tourism associations and other relevant stakeholders, in order to establish Namibia as a country at the forefront of DST.

In summary, astronomy provides a unique opportunity to build human capacity and diversify income generation in remote areas. Equally, dark sky tourism is well aligned with the 21st century ethos that tourism needs to be ecofriendly and sustainable. DST comes with many benefits, such as an increased awareness and understanding of science, environmental conservation (e.g. light pollution), and the celebration and preservation of indigenous heritage. These can all help to open up new avenues toward more meaningful and sustainable tourism practices.

\textbf{Acknowledgements.} This project is supported by the UKRI STFC Global Challenges Research Fund project ST/S002952/1 and Exeter College, Oxford.


\end{document}